\documentclass{eas}
\usepackage{graphicx}
\usepackage{natbib}
\newcommand{\kms}{km\,s$^{-1}$}

\newcommand{\msun}{M$_{\odot}$}

\newcommand{\fbin}{$f_\mathrm{bin}$}

\runningtitle{The VLT-Flames Tarantula Survey}
\begin{document}

\title{The VLT-Flames Tarantula Survey: an overview of the VFTS results so far}
\author{Hugues Sana}\address{Space Telescope Science Institute, 3700 San Martin Drive, Baltimore MD 21210, USA}
\author{the VLT-Flames Tarantula consortium}\address{The consortium members are as follows : Evans (PI), Bastian, Bestenlehner, Bonanos, Bressert, Brott, Cantiello, Clark, Crowther, de Koter, de Mink, Doran, Dufton, Dunstall, Garcia, Gieles, Graefener, H\'enault-Brunet, Herrero, Howarth, Izzard, Kennedy, K\"{o}hler, Langer, Lau, Lennon, Ma\'iz Apell\'aniz, Markova, McEnvoy, Najarro, Puls, Ram\'irez-Agudelo, Sab\'in-Sanjuli\'an, Sana, Schneider, Sim\'on-Di\'az, Smartt, Stroud, Taylor, van Loon, Vink, Walborn}
%
%
\begin{abstract}
The VLT-FLAMES Tarantula Survey (VFTS) has acquired multi-epoch spectroscopy of over 800 O, B and Wolf-Rayet stars in the 30 Doradus region with the aim to investigate a number of important questions related to the evolution of massive stars and of cluster dynamics. In this paper, I first provide an overview of the scientific results obtained by the VFTS consortium so far. I then review the constraints obtained on the multiplicity properties of massive stars in 30~Dor and compare them to our recent results from a Milky Way sample.
\end{abstract}
\maketitle
\section{Introduction}

The 30~Doradus (30~Dor) star forming region offers the largest concentration of massive star in the Local Group. Located in the Large Magellanic Clouds, most massive stars can still be resolved from the ground, offering the opportunity to individually study the components of what is our closest view of a massive starbursting region.

\begin{table}[t]
\caption{VFTS observational campaign; $R$ is the spectral resolving power.}
\centering
\label{tab: setup}
\begin{tabular}{llcccr}
\hline
Instr.  & Setting & Exp.time/OB & \# of epochs & $\lambda$-coverage & $R$ \\
      &         & [s]         &              & [\AA] \\
\hline
{\sc medusa} & LR02  & 2$\times$1815 & 6 & 3960-4564 & 7\,000 \\
{\sc medusa} & LR03  & 2$\times$1815 & 3 & 4499-5071 & 8\,500 \\
{\sc medusa} & HR15N & 2$\times$2265 & 2 & 6442-6817 &16\,000 \\
{\sc argus}    & LR02  & 2$\times$1815 & 5 & 3960-4570 &10\,500 \\
{\sc uves}     & 520   & 2$\times$1815 & 6 & 4175-6200 &53\,000 \\
\hline
\end{tabular}
\end{table}

\section{The VFTS observational campaign}
The VFTS data set has been collected in the course of a 160h large program (ID: 182.D-0222) at the European Southern Observatory. The observational campaign and target selection is extensively described in \citet{ETHB11_mnras}. We provide here a brief overview of the main properties of the campaign. The bulk of the data set has been obtained using the {\sc giraffe} spectrograph fed by the 132 {\sc medusa} fibers of {\sc flames}. 1000 different targets, brighter than 17~$mag$ in the $V$-band were observed using nine  plate configurations and three different observational setups (see Table~\ref{tab: setup}). 

Because of crowdedness, the central 1' was avoided in the main component of the survey. As a second component of the survey, the {\sc argus} integral-field unit, also feeding the {\sc giraffe} spectrograph, allowed us to study an additional 28 O stars and some WR stars in the densest parts of the field, exploring the suburbs of the massive star cluster R136 and providing crucial data for dynamical studies of the region. Last, yellow-red {\sc uves} spectra of a small sample of stars were obtained parallel to the {\sc argus} observations. In total over 22,000 individual spectra were acquired. A census of all massive stars in 30~Dor indicates that the VFTS has observed about 70\%\ of all massive stars in the region \citep{DCdK13_mnras}, but for the central cluster R136.

\begin{table}[t]
\caption{Overview of the VFTS scientific harvest}
\label{tab: sci}
\centering
\begin{tabular}{ll}
\hline
  Reference & Title \\
\hline
\citet{EWC10_mnras}  &  A Massive Runaway Star from 30 Doradus\\
\citet{ETHB11_mnras} &  I.~Introduction and observational overview  \\
\citet{TES11_mnras}  &  II.~R139 revealed as a massive binary system \\
\citeauthor{BVG11_mnras}   & III.~A very massive star in apparent isolation \\
\hspace*{2.5cm}(\citeyear{BVG11_mnras})  &   \multicolumn{1}{r}{from the massive cluster R136} \\
\citet{DDE11_mnras}  &  The Fastest Rotating O-type Star and  \\            
                  \multicolumn{2}{r}{Shortest Period LMC Pulsar -- Remnants of a Supernova  Disrupted Binary?}\\
\citet{BBE12_mnras}  &  IV.~Candidates for isolated high-mass star \\
                  \multicolumn{2}{r}{formation in 30 Doradus} \\
\citet{DFC12_mnras}  &  V.~The peculiar B[e]-like supergiant, VFTS~698,\\
                   \multicolumn{2}{r}{in 30 Doradus} \\
\citeauthor{HBGE12_mnras} &  VI.~Evidence for rotation   \\
\hspace*{2.5cm}(\citeyear{HBGE12_mnras})&  \multicolumn{1}{r}{of the young massive cluster R136} \\
\citeauthor{HBES12_mnras} &  VII.~A low velocity dispersion  \\
\hspace*{2.5cm}(\citeyear{HBES12_mnras}) & \multicolumn{1}{r}{for the young massive cluster R136} \\
\citet{SdKdM13_mnras}&  VIII.~Multiplicity properties \\
                   \multicolumn{2}{r}{of the O-type star population} \\
\citet{vLBT13_mnras} &  IX.~The interstellar medium seen  through diffuse \\ 
                   \multicolumn{2}{r}{interstellar bands and neutral sodium} \\
\citet{DLD13_mnras}  &  X.~Evidence for a bimodal distribution\\
                   \multicolumn{2}{r}{of rotational velocities for the single early B-type stars} \\
\citet{DCdK13_mnras} &  XI.~A census of the hot luminous stars  \\
                   \multicolumn{2}{r}{and their feedback in 30 Doradus} \\
\citeauthor{RASDS13}&  XII.~Rotational velocities of the single O-type stars\\
\hspace*{2.5cm}(\citeyear{RASDS13}) &\\
\citeauthor{SSSDH14}&  XIII.~On the nature of O Vz stars in 30 Doradus \\
\citeauthor{WSSD14} &  XIV.~The O-type stellar content of 30 Doradus \\
\citeauthor{JMA14} & XV. The extinction law in 30 Doradus\\
\hline
\end{tabular}
\end{table}

\section{The scientific harvest} 
The VFTS data set is very rich. The data reduction and analysis are described in an ongoing series of papers addressing a number of scientific questions related to the dynamics of the region, the properties of its massive stars, either seen as individual objects or as entire populations, and of the interstellar medium (Table~\ref{tab: sci}). In the present overview, we restrict ourselves to results related to the stellar components of 30~Dor.

\subsection{Individual objects of exception}

The first papers in the VFTS series mostly emphasize individual objects of exception:
\begin{enumerate}
\item[-] VFTS~016, an O2~III-If* star and one of the most massive runaway stars known so far, with a peculiar radial velocity (RV) of $-80$~\kms\ with respect to the systemic velocity of the region \citep{EWC10_mnras};
\item[-] R139, the most massive pair of evolved O stars, with O6.5 Iafc and O6 Iaf components and respective masses of about 80 and 65~\msun\ \citep{TES11_mnras};
\item[-] a number of massive stars found in apparent isolation from the central region \citep{BBE12_mnras}, including VFTS~682, a very massive star with a present day mass  of about 150~\msun\ \citep{BVG11_mnras}  and a spectrum almost identical to that of R136a3 in the core of the central cluster \citep{CSH10}. Some of these stars may be slow runaways. If formed in-situ, their presence outside the main massive star associations in the 30~Dor complex challenges the view that massive stars all have to be formed at the centre of massive stellar clusters and favours a more hierarchical view of star formation in giant molecular clouds;
\item[-] the possible association of VFTS~102, one of the two extremely fast rotators discovered in the VFTS \citep[$v_e \sin i \approx 610$~\kms, the other one being VFTS~285; see e.g.][]{RASDS13}, with the shortest period LMC pulsar \citep{DDE11_mnras};
\item[-] VFTS~698, a peculiar B[e]-like supergiant \citep{DFC12_mnras}.
\end{enumerate}

\subsection{The dynamics of the central region}

In two complementary papers, \citeauthor{HBGE12_mnras} (\citeyear{HBGE12_mnras}; \citeyear{HBES12_mnras}) investigated the dynamical properties of the inner 20~pc around R136. After rejecting the detected spectroscopic binaries and correcting for observational biases, \citet{HBES12_mnras} showed that the one-dimensional velocity dispersion was as low as 4 to 5~\kms, a value compatible with the R136 cluster being in virial equilibrium. \citet{HBGE12_mnras} further presented evidence of cluster rotation, possibly linked to a recent/ongoing merging within the core of 30~Dor \citep{SLG12}.

\subsection{Properties of the massive star populations}
Taking full advantage of the VFTS sample size, the remaining five papers published so far, and a number of others in preparation, investigate various aspects of the 30~Dor massive star populations and provide some of the best statistical constraints on massive star properties.  We present a brief overview of our results below. The multiplicity properties of the massive stars in 30~Dor will be separately discussed in the next section.

\begin{enumerate}
\item[-] \citeauthor{WSSD14} (to be submitted) present a detailed spectral classification of  the VFTS O-star sample and discuss the O-type stellar content of 30~Dor, illustrating the richness of both the VFTS data set and of the 30~Dor massive star population;
\item[-] \citet{DCdK13_mnras} performed a comprehensive census of hot luminous stars within a 10'-radius (about 150 pc) of R136 and estimated both the integrated ionising luminosity and stellar wind kinetic energy. These values were used to re-assess the star formation rate (SFR) of the region and determine the ionising photon escape fraction. The paper shows that, among the 1145 stars in the census, the 31 WR stars account for 40\%\ of the integrated ionising luminosity and for half the wind kinetic energy.  Stars with initial masses above 100~\msun\ (mostly H-rich WN stars) account for about one quarter of the global feedback, illustrating the need to take into account these very massive stars in popular stellar synthesis codes;
\item[-] Studying over 300 non supergiants early-B stars, \citet{DLD13_mnras} measured projected rotational velocities  up to about 450~\kms\ and revealed a bi-modal rotational velocity distribution. The bi-modal structure and its low velocity peak are in qualitative agreement with expectations from magnetic braking, though one can so far not exclude alternative scenarios;
\item[-] \citet{RASDS13} analysed 216 constant or weakly variable O stars. The O star rotational velocity distribution is dominated by a large population of slowly and moderately rotating O-type single stars ($v_e$ peaks at about 90~\kms), which represents about 75\%\ of the sample. The remaining 25\%\ of the sample populate a high-velocity tail that extends almost continuously from 200~to 600~\kms. The presence and magnitude of the high-velocity tail is in qualitative agreement with recent numerical simulations that investigate the effects of binary interactions through tides, mass-transfer and mergers \citep{dMLI13}. The high-velocity tail may thus be dominated by post-interaction binary products;
\item[-] \citeauthor{SSSDH14} (submitted) present atmosphere fitting analysis of a sample of about 80 VFTS O~V and O~Vz stars, shedding new light on the origin of the Vz phenomenon and concluding that the O~Vz stars mostly appear to be on, or very close to, the ZAMS. Some examples are however identified for which the Vz classification does not necessarily imply extreme youth. 
\end{enumerate}

\begin{table}[t]
\caption{Overview of the multiplicity properties of O and B stars in the VFTS and in a reference Milky Way (MW) sample}
\label{tab: mult}
\centering
\begin{tabular}{cccccc}
\hline
Quantity  & Function & Param. & VFTS O stars & VFTS B stars & MW O stars \\
\hline
$f_\mathrm{bin}$  & --            & $f_\mathrm{bin}$  & $0.51 \pm 0.04$ & $0.56\pm 0.11$ & $0.69\pm 0.09$ \\
$P$ (d)      & $(\log P)^\pi$ & $\pi$            & $-0.45 \pm 0.3$ & $-0.1\pm 0.5$  & $-0.55\pm 0.22$ \\
$M_2/M_1$       & $(M_2/M_1)^\kappa$      & $\kappa$         & $-1.0 \pm 0.4$  & $-2.5\pm0.8$   & $-0.10\pm 0.58$ \\
$e$             & $e^\eta$       & $\eta$           & $-0.5$ (fixed)  & $-0.5$ (fixed) & $-0.45\pm 0.17$ \\
\hline
\end{tabular}
\end{table}

\section{The multiplicity properties of the OB stars}

The multiplicity properties of the O- and early B-type star populations are investigated in \citet{SdKdM13_mnras} and Dunstall et al. (in prep.), respectively. The results for the B-type population mentioned below should be considered preliminary.  The spectroscopic binaries in the VFTS sample were identified through the presence of significant and relatively large RV variations: 46\%\ and 35\%\ of the O- and B-type samples present significant RV variations; 35\%\ and 26\%\ of the samples present RV variations large enough (larger than 20 and 16~\kms, respectively) to be considered as genuine spectroscopic binaries. To correct for observational biases and to retrieve the intrinsic distributions of binary parameters, we used a Monte Carlo approach. Adopting intrinsic period ($P$) and mass-ratio ($M_2/M_1$) distributions and an intrinsic spectroscopic binary fraction ($f_\mathrm{bin}$), we attempted to reproduce the properties of the VFTS observations in three aspects:
 \begin{enumerate}
\item[-] the observed binary fraction, i.e.\ the number of stars presenting significant RV variations larger than our adopted threshold;
\item[-] the amplitude of the RV signal, i.e.\ the maximum RV difference  between any pair of epochs that show significant variations larger that our threshold; 
\item[-] the timescale of the RV signal, i.e.\ the minimum separation between any pair of epochs that show significant variations larger that our threshold.
\end{enumerate}

The best-fit results are given in Table~\ref{tab: mult}. The multiplicity properties of the O- and B-type VFTS sample agree within error bars, but for the indexes of the mass-ratio distribution. We note however that our method is only weakly sensitive to the mass-ratio distribution exponent and one should restrain from drawing too firm conclusions in this regard.

In Table~\ref{tab: mult}, we further compare the VFTS results to the Milky Way sample of young stellar clusters of \citet{SdMdK12}. While there are other recent surveys on the multiplicity of massive stars \citep[e.g.][see also \citeauthor{SaE11}, \citeyear{SaE11}, and references therein]{MHG09, KiK12, CHN12}, the   MW sample has been corrected for observational biases using the same underlying hypotheses and a very similar methodology, decreasing the risk of systematic error in the comparison. Given the relatively low densities and  young ages of the stellar clusters considered, the results of \citet{SdMdK12} provide one of the best available estimates for the initial binary properties in 5,000-10,000~\msun\ clusters. 

Interestingly, the orbital period distribution is very similar while the binary fraction is lower in 30~Dor by about 18\%. This raises the question whether the initial binary fraction was different in 30~Dor compared to the MW clusters, possibly reflecting the different environmental conditions, or whether 30~Dor started with similar initial conditions but already destroyed some of its original binary population through dynamical interaction or binary evolution. 

Building upon the results of \citet{dMLI13} mentioned earlier, \citeauthor{dMSL14} (subm.) estimated, among other things, the overall detection probability of post-binary interaction products through (idealized) RV campaigns. Assuming an initial binary fraction of 70\%\ and continuous star formation, \citet{dMSL14} (subm.) computed that a spectroscopic binary of only about 55\%\ would be observed, corresponding to binaries with peak-to-peak RV variations in excess of 20~\kms\ (assuming circular obit and considering the primary stars only). While numerical simulations including the actual star formation history of the 30~Dor complex would be required to draw definite conclusions, the  similarity between the numerical predictions and the actual VFTS observations suggest that binary evolution may play a significant role in explaining the different \fbin\ values measured in 30~Dor and the MW clusters.

\section{The future}
Over the last few years, the VLT-Flames Tarantula survey has provided state-of-the art new observational constraints on the stellar content and the dynamics of the 30~Dor region. A number of VFTS studies are still ongoing, that will address further the gas and extinction properties, the dynamics of the region, the physical properties and evolutionary status of the individual stars, as well as detailed quantitative comparison to evolutionary models. Monitoring of many of the newly discovered spectroscopic binaries has further been initiated (PI: Sana) and will provide orbital parameters and, in many cases, disentangled blue spectra of the individual components for atmosphere analysis.

 The 30~Dor region is also at the heart of several HST programs to obtain spatially resolve spectroscopy of the R136 cluster \citep{BWC13},  panchromatic photometry of its stellar populations \citep{SAL13_mnras} and high accuracy proper motions \citep{dMLS12_mnras}. Including existing {\sc spitzer} data \citep{wbs13} and a future  {\sc chandra} visionary program (T-Rex; PI: Townsley), these space-based data will critically complement the ground-based results obtained by the VFTS consortium so far. The final aim is to achieve an as complete as possible multi-wavelength view of the closest massive starburst, of its gas phase and its stellar population to better understand massive starbursting regions in the distant universe.


\bibliographystyle{astron}
\bibliography{/home/hsana/Dropbox/literature}

\begin{thebibliography}{}

\bibitem[\protect\astroncite{{Bestenlehner}  {et~al.}}{2011}]{BVG11_mnras}
{Bestenlehner}, J.~M.  {et~al.}: 2011,
\newblock {\em \aap} {\bf 530}, L14

\bibitem[\protect\astroncite{{Bostroem} et~al.}{2013}]{BWC13}
{Bostroem}, K., {Walborn}, N., {Crowther}, P., {Caballero-Nieves}, S.,
  {Lennon}, D., and {Ma{\'{\i}}z Apell{\'a}niz}, J.: 2013,
\newblock in {\em Massive Stars: From alpha to Omega}

\bibitem[\protect\astroncite{{Bressert}  {et al.}}{2012}]{BBE12_mnras}
{Bressert}, E.  {et al.}: 2012,
\newblock {\em \aap} {\bf 542}, A49

\bibitem[\protect\astroncite{{Chini} et~al.}{2012}]{CHN12}
{Chini}, R., {Hoffmeister}, V.~H., {Nasseri}, A., {Stahl}, O., and {Zinnecker},
  H.: 2012,
\newblock {\em \mnras} {\bf 424}, 1925

\bibitem[\protect\astroncite{{Crowther} et~al.}{2010}]{CSH10}
{Crowther}, P.~A., {Schnurr}, O., {Hirschi}, R., {Yusof}, N., {Parker}, R.~J.,
  {Goodwin}, S.~P., and {Kassim}, H.~A.: 2010,
\newblock {\em \mnras} {\bf 408}, 731

\bibitem[\protect\astroncite{{de Mink}  et~al.}{2012}]{dMLS12_mnras}
{de Mink}, S.~E.  et~al.: 2012,
\newblock in {\em American Astronomical Society Meeting Abstracts}, Vol. 219 of
  {\em American Astronomical Society Meeting Abstracts}, p. 151.13

\bibitem[\protect\astroncite{{de Mink} et~al.}{2013}]{dMLI13}
{de Mink}, S.~E., {Langer}, N., {Izzard}, R.~G., {Sana}, H., and {de Koter},
  A.: 2013,
\newblock {\em \apj} {\bf 764}, 166

\bibitem[\protect\astroncite{{de Mink} et~al.}{}]{dMSL14}
{de Mink}, S.~E., {Sana}, H., {Langer}, N., {Izzard}, R.~G., and {Schneider},
  F.,
\newblock {\em \apj} submitted

\bibitem[\protect\astroncite{{Doran}  et~al.}{2013}]{DCdK13_mnras}
{Doran}, E.~I.  et~al.: 2013,
\newblock {\em \aap} {\bf 558}, A134

\bibitem[\protect\astroncite{{Dufton}  et~al.}{2011}]{DDE11_mnras}
{Dufton}, P.~L.  et~al.: 2011,
\newblock {\em \apjl} {\bf 743}, L22

\bibitem[\protect\astroncite{{Dufton}  et~al.}{2013}]{DLD13_mnras}
{Dufton}, P.~L.  et~al.: 2013,
\newblock {\em \aap} {\bf 550}, A109

\bibitem[\protect\astroncite{{Dunstall}  et~al.}{2012}]{DFC12_mnras}
{Dunstall}, P.~R.  et~al.: 2012,
\newblock {\em \aap} {\bf 542}, A50

\bibitem[\protect\astroncite{{Evans}  et~al.}{2010}]{EWC10_mnras}
{Evans}, C.~J.  et~al.: 2010,
\newblock {\em \apjl} {\bf 715}, L74

\bibitem[\protect\astroncite{{Evans}  et~al.}{2011}]{ETHB11_mnras}
{Evans}, C.~J.  et~al.: 2011,
\newblock {\em \aap} {\bf 530}, A108

\bibitem[\protect\astroncite{{H{\'e}nault-Brunet} and {et
  al.}}{2012a}]{HBGE12_mnras}
{H{\'e}nault-Brunet}, V.  et~al.: 2012a,
\newblock {\em \aap} {\bf 545}, L1

\bibitem[\protect\astroncite{{H{\'e}nault-Brunet} and {et
  al.}}{2012b}]{HBES12_mnras}
{H{\'e}nault-Brunet}, V.  et~al.: 2012b,
\newblock {\em \aap} {\bf 546}, A73

\bibitem[\protect\astroncite{{Kiminki} and {Kobulnicky}}{2012}]{KiK12}
{Kiminki}, D.~C. and {Kobulnicky}, H.~A.: 2012,
\newblock {\em \apj} {\bf 751}, 4

\bibitem[\protect\astroncite{{Ma{\'{\i}}z Apell{\'a}niz} and {et
  al.}}{}]{JMA14}
{Ma{\'{\i}}z Apell{\'a}niz}, J.  et~al.,
\newblock {\em \aap} in preparation

\bibitem[\protect\astroncite{{Mason} et~al.}{2009}]{MHG09}
{Mason}, B.~D., {Hartkopf}, W.~I., {Gies}, D.~R., {Henry}, T.~J., and {Helsel},
  J.~W.: 2009,
\newblock {\em \aj} {\bf 137}, 3358

\bibitem[\protect\astroncite{{Ram{\'{\i}}rez-Agudelo} and {et
  al.}}{2013}]{RASDS13}
{Ram{\'{\i}}rez-Agudelo}, O.~H.  et~al.: 2013,
\newblock {\em \aap} in press (arXiv: 1309.2929)

\bibitem[\protect\astroncite{{Sabbi} et~al.}{2013}]{SAL13_mnras}
{Sabbi}, E., et~al.: 2013,
\newblock {\em \aj} {\bf 146}, 53

\bibitem[\protect\astroncite{{Sabbi} et~al.}{2012}]{SLG12}
{Sabbi}, E., {Lennon}, D.~J., {Gieles}, M., {de Mink}, S.~E., {Walborn}, N.~R.,
  {Anderson}, J., {Bellini}, A., {Panagia}, N., {van der Marel}, R., and
  {Ma{\'{\i}}z Apell{\'a}niz}, J.: 2012,
\newblock {\em \apjl} {\bf 754}, L37

\bibitem[\protect\astroncite{{Sab{\'{\i}}n-Sanjuli{\'an}} and {et
  al.}}{}]{SSSDH14}
{Sab{\'{\i}}n-Sanjuli{\'an}}, C.  et~al.,
\newblock {\em \aap} submitted

\bibitem[\protect\astroncite{{Sana} et~al.}{2012}]{SdMdK12}
{Sana}, H., {de Mink}, S.~E., {de Koter}, A., {Langer}, N., {Evans}, C.~J.,
  {Gieles}, M., {Gosset}, E., {Izzard}, R.~G., {Le Bouquin}, J.-B., and
  {Schneider}, F.: 2012,
\newblock {\em Science} {\bf 337}, 444

\bibitem[\protect\astroncite{{Sana}  et~al.}{2013}]{SdKdM13_mnras}
{Sana}, H.  et~al.: 2013,
\newblock {\em \aap} {\bf 550}, A107

\bibitem[\protect\astroncite{{Sana} and {Evans}}{2011}]{SaE11}
{Sana}, H. and {Evans}, C.~J.: 2011,
\newblock in {C.~Neiner, G.~Wade, G.~Meynet, \& G.~Peters} (ed.), {\em IAU
  Symposium}, Vol. 272 of {\em IAU Symposium}, pp 474--485

\bibitem[\protect\astroncite{{Taylor}  et~al.}{2011}]{TES11_mnras}
{Taylor}, W.~D.  et~al.: 2011,
\newblock {\em \aap} {\bf 530}, L10

\bibitem[\protect\astroncite{{van Loon}  et~al.}{2013}]{vLBT13_mnras}
{van Loon}, J.~T.  et~al.: 2013,
\newblock {\em \aap} {\bf 550}, A108

\bibitem[\protect\astroncite{{Walborn} et~al.}{2013}]{wbs13}
{Walborn}, N.~R., {Barb\'a}, R., and {Sewilo}, M.~M.: 2013,
\newblock {\em \aj} in press (arXiv:1302.3533)

\bibitem[\protect\astroncite{{Walborn}  et~al.}{}]{WSSD14}
{Walborn}, N.~R.  et~al.,
\newblock {\em \aap} in prep.

\end{thebibliography}

\end{document}